# Density of Phonon States in Superconducting FeSe as a Function of Temperature and Pressure

V. Ksenofontov,[1] G. Wortmann,[2] A.I. Chumakov,[3] T. Gasi,[1] S. Medvedev,[1,4]
T.M. McQueen,[5§] R.J. Cava,[5] and C. Felser[1]

[1]*Institut für Anorganische und Analytische Chemie, Johannes Gutenberg-Universität, D-55099 Mainz, Germany*
[2]*Department Physik, Universität Paderborn, D-33095 Paderborn, Germany*
[3]*European Synchrotron Radiation Facility, BP220, F-38043 Grenoble Cedex, France*
[4]*Max-Planck-Institute for Chemistry, D-55128 Mainz, Germany*
[5]*Department of Chemistry, Princeton University, Princeton NJ 08544, USA*

**Abstract:**

The temperature and pressure dependence of the partial density of phonon states of iron atoms in superconducting $Fe_{1.01}Se$ was studied by $^{57}Fe$ nuclear inelastic scattering (NIS). The high energy resolution allows for a detailed observation of spectral properties. A sharpening of the optical phonon modes and shift of all spectral features towards higher energies by ~4% with decreasing temperature from 296 K to 10 K was found. However, no detectable change at the tetragonal – orthorhombic phase transition around 100 K was observed. Application of a pressure of 6.7 GPa, connected with an increase of the superconducting temperature from 8 K to 34 K, results in an increase of the optical phonon mode energies at 296 K by ~12%, and an even more pronounced increase for the lowest-lying transversal acoustic mode. Despite these strong pressure-induced modifications of the phonon-DOS we conclude that the pronounced increase of $T_c$ in $Fe_{1.01}Se$ with pressure cannot be described in the framework of classical electron-phonon coupling. This result suggests the importance of spin fluctuations to the observed superconductivity.

# Introduction

The recent discovery of superconductivity with $T_c$ values up to 56 K in FeAs-based layered compounds has attracted tremendous interest [1-6] and has also stimulated the search for further related superconductors. The observation of superconductivity in the $\beta$-Fe$_{1+x}$Se (x >0) system [7,8] with $T_c \cong 8.5$ K at ambient pressure together with the dramatic increase of $T_c$ with pressure up to 37 K [9-11] provides a "simple" model compound [12] since $\beta$-FeSe with the PbO-type tetragonal structure is built up of layers of FeSe$_4$ edge-sharing tetrahedra, the same structural motif as in the FeAs-based superconductors. The superconducting (sc) properties of FeSe depend strongly on the stoichiometry [8] and the structure [10,13], as demonstrated by the fact that sc Fe$_{1.01}$Se undergoes a continuous tetragonal to orthorhombic transition (T-O) around 100 K, whereas the non-sc Fe$_{1.03}$Se remains tetragonal by x-ray diffraction [13]. One can suppose that the structural instability in the sc phase has relations to the mechanism of superconductivity. Important in this context was the finding that the T-O phase transition in sc FeSe systems is not magnetically driven [8,9,13].

Essential progress was achieved in high-pressure studies of FeSe systems, establishing the structural phase diagram and border-lines of superconductivity [9,10]. In particular, i) an enormous increase of $T_c$ attaining 37 K was found in Fe$_{1.01}$Se under pressure of ~9.0 GPa; ii) subsequent decrease of $T_c$ accompanied by the structural phase transformation into the semiconducting high-pressure phase was revealed for p > 9.0 GPa, and iii) non-magnetic ground states were found at 4.2 K for both the sc orthorhombic and non-sc high-pressure phases by means of $^{57}$Fe-Mössbauer spectroscopy [9].

The origin of Cooper pairing in these new superconductors is the most important question. Replacement of classical electron-phonon interaction by a mechanism based on spin fluctuations and/or spin-density waves is considered as an alternative microscopic mechanism of superconductivity in Fe-based pnictides [14]. Indeed, there is experimental evidence for antiferromagnetic fluctuations in FeSe observed with $^{77}$Se-NMR, enhanced under pressure in the orthorhombic low-temperature phase [15]. These spin fluctuations should originate from the conduction band, since $^{57}$Fe-Mössbauer investigations at ambient and high pressure [8,9] as well as in external applied magnetic fields [16-18] did not observe magnetic order or local Fe moments, respectively.

The observation of a large iron isotope effect, here an increase of $T_c$ for the lighter $^{54}$Fe isotope, in the related FeAs-systems Ba$_{1-x}$K$_x$Fe$_2$As$_2$ and SmFeAsO$_{1-x}$F$_x$, suggests an important role of Fe phonon modes for the observed superconductivity [19]. Very recently, a similar isotope effect has been reported for the present FeSe system [20], demonstrating again that the vibrations of iron atoms are involved in the sc mechanism, possibly connected, as discussed in [14,15], with spin fluctuations. Therefore a detailed study of the partial density of phonon states (phonon DOS) of Fe atoms, which are here the active "players", is of actual interest. Taking into account the enormous increase of $T_c$ in $Fe_{1.01}Se$ under pressure [9], the study of the Fe-partial phonon DOS under pressure is of utmost importance.

In this paper, we report the results of $^{57}Fe$ nuclear inelastic scattering (NIS) spectroscopy performed on $Fe_{1.01}Se$ with a high energy resolution of 0.75 meV, delivering the Fe-partial phonon-DOS as a function of temperature and pressure, yielding important information about the various acoustic and optic modes. The application of a pressure of 6.7 GPa, where the present $Fe_{1.01}Se$ sample exhibits a $T_c$ value of 34 K, just below the maximum $T_c$ of 37 K, and remains in the pure PbO-phase, leads to a strong increase in energy of the phonon spectrum, being most pronounced for the low-lying transversal acoustic modes.

The paper is organized as follows. After the experimental details, we present the typical $^{57}Fe$-NIS spectra and the derived Fe partial phonon-DOS measured as a function of temperature and pressure. The present Fe-partial phonon-DOS is compared with a neutron scattering study of FeSe which gives information on the total phonon-DOS [21]. Phonon-modes involving vibrations of both Fe and Se as well as exclusively either Fe or Se can be distinguished with the help of theoretical studies [22,23]. Informative for the present sc FeSe-system, called in literature "11 system" [12], is the comparison with $^{57}Fe$-NIS and Raman studies of other Fe-based sc systems, e.g. the 1111 system $LaFeAsO_{1-x}F_x$ [24] as well as the 111 system $Ba_{1-x}K_xFe_2As_2$ [25]. Finally we discuss the changes in the phonon-DOS under pressure with the strong increase of $T_c$ in the framework of the theoretical approaches for conventional superconductivity.

**Experimental**

A sample of sc $Fe_{1.01}Se$, enriched with 20% $^{57}Fe$, was prepared in the same way as previously reported [8]. Laboratory XRD characterization of the present $Fe_{1.01}Se$ sample confirmed the tetragonal PbO-type structure at ambient pressure. The sample is the same as the one used for the $^{57}Fe$-Mössbauer studies described in Refs [9,13]. Storage and handling of the sample was done under strictly inert conditions [8,26]. For the high-pressure (hp) study a Paderborn-type diamond anvil cell [27,28] was used, with a diamond anvil flat of 500 μm and a Be gasket with a sample hole diameter of 200 μm. Silicon oil was used as inert pressure transmitting medium. The pressure was determined by the ruby fluorescence method. The NIS experiments were performed at the Nuclear Resonance beamline ID18 [29] of the European Synchrotron Radiation Facility (ESRF), operated in the 16-bunch mode of the storage ring with a nominal electron current of 90 mA. A four-bounce in-line [30] high-resolution monochromator, consisting of two pairs of Si(4 0 0) and Si (12 2 2) crystals, provided the flux of $4\times10^9$ photons per second within the energy bandwidth of 0.75 meV (FWHM) at the $^{57}Fe$ resonance energy of 14.413 keV [31]. Energy calibration was performed using the dominant [57]Fe stretching mode at 74.0 meV in $(NH_4)_2Mg^{57}Fe(CN)_6$ [32]. For the hp experiment, the monochromatized SR beam was focussed on a spot of $5 \times 13$ μm$^2$, so fully hitting the pressurized FeSe sample. [57]Fe-NIS spectra were recorded in an energy range ±80 meV with a stepwidth of 0.20 meV. For each temperature, about 20 NIS spectra, each taking about 40 min, were added. For the hp study, 30 NIS spectra were added, which had less counting rate due to the reduced geometry and strong absorption for the scattered 14.413 keV gamma and Fe $K_{\alpha,\beta}$ x-rays by the hp cell [28]. Evaluation of the Fe partial phonon-DOS from the NIS spectra is described in [32-34]. The derived phonon-DOS, presented below, are smoothed by a linear filter function with a width of the averaging window equal to the experimental instrumental resolution of 0.75 meV.

**Results and Discussion**

Two selected spectra of nuclear inelastic scattering at the [57]Fe resonance in $Fe_{1.01}Se$, measured at T = 296 K and T = 10 K, shown in Fig. 1, exhibit the typical temperature dependence of the inelastic Stokes and anti-Stokes side wings originating from the creation and annihilation of phonons, respectively [32-34]. From such NIS-spectra, measured at various temperatures in the tetragonal (296 K and 110 K) and in the orthorhombic phase (66 K and 10 K), the derived Fe-partial phonon-DOS, $g(E)$, are shown in Fig. 2. This figure also presents the Fe phonon-DOS measured at T = 296 K under a pressure of 6.7 GPa. The high energy resolution over the whole spectral range allows an assignment of characteristic acoustic and optical branches. With decreasing temperatures a shift of all characteristic features, called here in short modes [35], to higher energies (by ~4%) and a sharpening especially of the optical modes was observed. The energies of eight pronounced features in the Fe-partial phonon-DOS, best resolved in the 10 K data and labeled (1)-(8) in Fig.2, and their dependencies on temperature and pressure are presented in the Table 1. We follow in our assignment of these spectral features a recent neutron scattering study of FeSe [21] as well as two theoretical calculations of the phonon dispersions and derived phonon-DOS in FeSe [22,23].

We start at the low energy part of the Fe-partial phonon-DOS at 10 K in Fig. 2, where we attribute two features at 5.6 and 9.4 meV, designated (1) and (2), to transversal acoustic (TA) modes, better resolved in the so-called reduced form of the DOS, $g(E)/E^2$, shown in Fig. 3. The broad, less structured features from 15 to 18 meV, labeled (3), are attributed to mostly longitudinal acoustic (LA) modes. This assignment of the acoustic modes is in principle agreement with the theoretical calculations [22,23], indicating an upper value for "pure" LA frequencies around 18 meV. The highly anisotropic layered structure as well as the large c-axis compressibility due to the soft, van der Waals-like, interlayer bonding, should be reflected in the acoustic modes in c-direction. In Ref. 22, such a soft TA mode was calculated around 2.5 meV for the Γ-Z branch of the Brillouin zone. We could not detect, in agreement with the neutron study [21], any structure at this low energy, still well resolvable with the present experimental resolution. We attribute, however, the mode (1) around 5.4 meV to this soft TA branch. This assignment is supported by the strong pressure induced shift of this mode as discussed below.

Above ~20 meV, we assign the well-resolved spectral features in the DOS to optical modes. These optical modes exhibit, due to the small energy variation in their dispersion relations and due to the high spectral resolution of the present study, well-resolved structures which can be adjusted by Voigt profiles as shown in the 10 K data of Fig. 2 [36].

By comparison with the theoretical and experimental phonon studies, the prominent peak at 25.5(3) meV, labeled (5) in Fig. 2, can be attributed to the Fe $B_{1g}$ Raman mode in the calculated phonon DOS [22,23]. $B_{1g}$ represents a special mode, involving mainly the displacements of Fe atoms vibrating in opposite directions parallel to the c-axis in the Fe-Se layers, see [37]. Most interestingly, this mode is also observed around 25 meV in $^{57}$Fe-NIS studies of the other Fe-based high-$T_c$ systems with Fe-As layers, here in LaFeAsO$_{1-x}$F$_x$ [24] and in Ba$_{1-x}$K$_x$Fe$_2$As$_2$ [25] and is characteristic to the Fe-X (X = Se, As) layers [23]. Raman studies of this mode, including its temperature dependence, are well documented [37,38], with a very similar dependence as observed here. Since the mass of the vibrating Fe atoms is approximately the same in these different studies [39], the relative small differences in the frequencies of this Raman mode must be attributed to differences in Fe-X bonding strength. Both theoretical calculations of the phonon-DOS in FeSe put this $B_{1g}$ Fe-mode to considerably higher energies around 29 meV [22] and around 31 meV [23]. Very helpful in this context is the decomposition of the theoretical phonon-DOS of Ref. 22 into the relative contributions of the Fe and Se constituents in the neutron study [21,40].

We proceed now to the well-resolved group of optical modes in the energy region from 30 to 40 meV, consisting of combined Fe and Se optical modes. The mode (8) with the highest energy of 38.7 meV is attributed to the $E_g^{(2)}$ mode from comparison with theoretical and other studies [21,37,38,41]. The dominant peak (6) at 31.6 meV can be attributed to the $E_u^{(2)}$ mode. Again, as for the $B_{1g}$ mode, the theoretical calculations predict higher values for these modes (6) and (8), calculated at 35 meV and 40 meV in Ref. 22 and to 37 meV and 42 meV in Ref. 23. Finally, we tentatively assign the relatively weak structure at 20.6 meV, best resolved at 10 K, to the $E_g^{(1)}$ mode.

By lowering the temperature from 296 K to 10 K, there is a sharpening of the spectral features in the phonon-DOS of FeSe, evident from Fig. 2, accompanied by a shift to higher energies by about 4% (see Tab. 1 and Ref 42). Comparison of the well-resolved phonon DOS at 110 K (tetragonal phase) and at 66 K (orthorhombic phase) reveal, beside small changes in energy due to the lowering of temperature, no difference which could be attributed to the subtle T-O phase transition of sc Fe$_{1.01}$Se around 100 K [13]. This finding is in principal agreement with the phonon DOS data derived in the neutron study with less spectral resolution [21]. It should be mentioned that

$^{57}$Fe-Mössbauer studies of the present sc Fe$_{1.01}$Se sample as well of a non-sc Fe$_{1.03}$Se sample [13,18] reveal no differences in the quadrupole splitting, but for the sc Fe$_{1.01}$Se sample a slightly reduced f-factor (by ~1.5 %) in the orthorhombic phase. This indicates that the Fe binding strength in the sc orthorhombic phase is slightly reduced in comparison to the non-sc Fe$_{1.03}$Se sample [18]. Such a behavior is also reflected in the mean-squared vibration amplitudes derived in the XRD study data of these samples [13].

The normalized Fe-partial phonon DOS allows us to derive elastic and thermodynamic parameters of the Fe sublattice, similar as done in a related $^{119}$Sn-NIS study of SnO [41], a compound with the same PbO-structure as FeSe. Table 2 presents Lamb-Mössbauer factor $f_{LM}$ and the mean force constant D, derived from the $^{57}$Fe-partial phonon-DOS of sc Fe$_{1.01}$Se at 296 K. Other elastic and thermodynamic parameters of the Fe sublattice from the present sc Fe$_{1.01}$Se sample as well as from a non-sc Fe$_{1.03}$Se sample derived from their phonon-DOS as well as from their Mössbauer spectra will be given in a forthcoming publication [18].

Fig. 2 (bottom) shows the phonon DOS of Fe$_{1.01}$Se measured at 296 K under a pressure of 6.7 GPa. At this pressure, the present sample exhibits superconductivity with T$_c$ = 34 K and is free from the high-pressure phase, which starts to appear above 7 GPa and increases in amount at higher pressures [9,10]. The phonon DOS of Fe$_{1.01}$Se at 6.7 GPa clearly reflects the strong increase of the spectral features towards higher energy due to the reduction of the molar volume by ~14% [10,43], evident for the optical modes (5)–(8) by a shift of about 12%. One observes even more significant modifications of the phonon-DOS $g(E)$, especially for the transverse acoustic modes, where mode (1) is shifted by ~30% from 5.4 meV to ~7 meV and mode (2) by ~14% from 8.8 meV to 10.0 meV. This behavior is exemplified in the low-energy range, where the Debye approximation is valid, in the plot of the so-called reduced phonon-DOS, $g(E)/E^2$, indicating by the extrapolation of the spectral features to $E = 0$, that the Debye sound velocity and therefore the bonding strength has considerably increased under pressure (Fig. 3). Calculation of the averaged Debye sound velocity $\langle v_D \rangle$ from acoustic vibrations was done using the relation [44,45]:

$$\lim_{E \to 0} \frac{g(E)}{E^2} = \alpha = \left( \frac{2m}{m+M} \right) \frac{1}{2\pi^2 \hbar^3 n \langle v_D \rangle^3} \qquad (1)$$

Here $m$ = 57 for Fe and $M$ = 79 for Se, $n = 4/V_0$ is the number of atoms per unit volume, $V_0$ is the volume of the unit cell. With $V_0 = 78.58 \cdot 10^{-30}$ m$^3$ (ambient pressure) and $V_0 = 67.6 \cdot 10^{-30}$ m$^3$ ($p = 6.7$ GPa), the derived $\langle v_D \rangle$ values are $2.05(4) \cdot 10^3$ m/s and $2.25(7) \cdot 10^3$ m/s at ambient pressure and $p = 6.7$ GPa, respectively. From the above $\alpha$ values, using the relation $\Theta_{D,LT} = (3/\alpha)^{1/3}/k_B$ [44,45], the corresponding low-temperature (LT) Debye temperatures $\Theta_{D,LT}$ = 240(5) K (ambient pressure) and $\Theta_{D,LT}$ = 278(8) K (6.7 GPa) were derived. They are listed with the values of $\alpha$ and $\langle v_D \rangle$ in Tab. 3.

These $\Theta_{D,LT}$ values reflect mainly the acoustic modes, they are smaller than the values $\Theta_D$ = 285(4) K and $\Theta_D$ = 317(6) K calculated at ambient pressure and $p$ = 6.7 GPa, respectively, from the corresponding Lamb-Mössbauer factors, $f_{LM}$ = 0.60(1) and 0.66(1), representing more the whole phonon-DOS. The present observation of rather low $\Theta_{D,LT}$ values in comparison to the $\Theta_D$ values is in accord with the observations in SnO [41] with a small bulk modulus, $B_0$ = 38 GPa, similar to $B_0$ = 31 GPa derived for FeSe [10]. These small values of $B_0$ originate mainly from the large c-axis compressibility. The present sound velocities are smaller than $<v_D>$ = 2.46·$10^3$ m/s derived in sc and non-sc $Ba_{1-x}K_xFe_2As_2$ from a $^{57}$Fe-NIS study [25].

The pressure-induced energy shift of certain well-resolved modes can be expressed by mode-specific Grüneisen parameter $\gamma_i$ according to the Debye-Grüneisen model by $\gamma_i = dlnE_i/dlnV$. The corresponding values of $\gamma_i$ derived for the 6.7 GPa data with $dlnV$ = 0.14 [10,43] are listed in Tab. 1. While the optic modes with $i$ = 5, 6, 8 exhibit $\gamma_i$ values from 0.7(1) to 0.9(1), the two lowest acoustic modes exhibit values of $\gamma_1$ = 2.4(3) and $\gamma_2$ = 1.0(1). As mentioned above, the large value for the mode (1) is attributed to the strong compression of the c-axis and the interlayer distance. The optical modes, on the other hand, experience a smaller change of the intra-layer Fe-Se distances due to the smaller compression of the a-axes than the c-axis, $dlnc/dp$ = 2.5 $dlna/dp$ [10]. Consequently, the pressure-induced increase in energy of the spectral features of the phonon-DOS in sc $Fe_{1.01}Se$ can be entirely explained by the reduction of the local metric at the Fe constituents. We do not observe any lattice instability, i.e. a decrease [25] in the energy of certain modes which may be relevant to the mechanism of the superconductivity, here, despite the strong increase of $T_c$ with pressure.

We discuss the strong increase of $T_c$ in $Fe_{1.01}Se$ under pressure using the McMillan formula for conventional, electron-phonon mediated superconductivity [46], following recent discussions on similar Fe-based superconductors [47, 48]:

$$T_C = \frac{\Theta_D}{1.45}\exp[-\frac{1.04(1+\lambda)}{\lambda-\mu^*(1+0.62\lambda)}] \qquad (2)$$

where $\Theta_D$ is the Debye temperature derived from the $f_{LM}$, $\mu^* \cong 0.1$ is the Coulomb pseudopotential, the electron-phonon coupling constant $\lambda$ is proportional to the average inverse square of the phonon frequency $\langle\omega^{-2}\rangle$. With $\Theta_D$ = 285(4) K one can find from (2) that $T_c \cong 8.5$ K observed at ambient pressure corresponds to $\lambda$ = 0.65(1). This value is between a theoretically derived value of $\lambda$ = 0.21 for LaOFeAs [48, 49] and a value $\lambda$ = 1.3 deduced for $PrFeAsO_{1-x}F_x$ from transport measurements [49, 50]. The present pressure-induced hardening of the average phonon mode energies by 12% or even the increase of TA soft mode energy by 30% results with the above equation (2) in an increase of $T_c$ = 8.5 K by only a few percent which is much lower than the experimental increase of $T_c$ to 34 K at 6.7 GPa [9]. Our conclusion is that $T_c$ estimation based on the McMillan formalism for superconductivity mediated by electron-phonon interaction [46] cannot explain the strong rise of $T_c$ in $Fe_{1.01}Se$ under pressure considering only modifications of the phonon spectrum. Very similar conclusions were obtained for the LaOFeAs system [48] using the Allen-Dynes formalism [51].

One factor which could be responsible for the observed $T_c$ enhancement in $Fe_{1.01}Se$ under pressure is a strong modification of chemical or valence state of the $Fe^{2+}$ ions, causing substantial changes at the electronic density of states at the Fermi level. However, the present experimental data does not seem to support this hypothesis. Such pressure-induced changes should be reflected in the Mössbauer isomer shift $\delta$ at the $^{57}Fe$ nuclei. Analysis of pressure dependence of $\delta$ in the tetragonal phase of $Fe_{1.01}Se$ [9] provides a small and almost linear variation of $\delta$ with a pressure, corresponding to a volume coefficient of $d\delta/dlnV$ = 0.40(2) mm/s, a value typical for the covalent (or semi-metallic) $Fe^{2+}$ state [16, 52], with no indications for any abnormal changes of the $s$-electron density at pressures up to 9 GPa [18].

Instead, from the previous studies of $Fe_{1.01}Se$, the observation of strong spin fluctuations in $^{77}Se$-NMR studies is most relevant for the sc mechanism [15]. These spin fluctuations, originating from d-electrons the conduction band, are enhanced in the sc orthorhombic phase and also, most importantly, strongly under pressure. This indicates that antiferromagnetic spin fluctuations, coupled to phonon modes, may play a decisive role in the superconductivity in $Fe_{1.01}Se$ [15, 22]. Of actual interest here the $FeSe_{1-x}Te_x$ systems [47, 53], exhibiting an increase of $T_c$ with Te substitution and with pressure. Investigation of the phonon-DOS in these systems might provide additional information about the respective mechanisms for superconductivity.

**Conclusion**

The temperature and pressure dependence of the partial density of phonon states of iron atoms in superconducting $Fe_{1.01}Se$ was studied by $^{57}Fe$-NIS. The high energy resolution allows for a detailed observation of spectral properties. A sharpening of the optical phonon modes and shift of all spectral features towards higher energies by ~4% with decreasing temperature from 296 K to 10 K, was seen. However, no detectable change at the tetragonal – orthorhombic phase transition around 100 K was observed. Application of a pressure of 6.7 GPa, connected with an increase of the superconducting temperature from 8.5 K to 34 K, resultsed in an increase of the phonon mode energies at 296 K by ~12%, and an even more pronounced increase for the transversal acoustic modes.

The present data on the partial Fe phonon-DOS with the unambiguous identification of various Raman as well as acoustic modes with high energy accuracy should initiate theoretical calculations with improved interaction parameters to meet the present experimental results. Such an improvement in the theoretical description may also provide new and better approaches to the description of the electronic and magnetic interactions, including spin-fluctuations [14].

Despite the pronounced pressure-induced modifications of the phonon-DOS we conclude that the strong increase of $T_c$ in $Fe_{1.01}Se$ with pressure cannot be described in the framework of classical electron-phonon coupling. This result suggests the importance of alternative mechanisms, e.g. spin fluctuations, to describe the observed superconductivity.

**Acknowledgements**

The work at Mainz was financially supported by the DFG in the Collaborative Research Center Condensed Matter Systems with Variable Many-Body Interactions (TRR 49). G.W. gratefully acknowledges support by the DFG grant WO209/13. The work at Princeton was supported primarily by the US Department of Energy, Division of Basic Energy Sciences, Grant No. DE-FG02-98ER45706. T.M.M. gratefully acknowledges the support of the National Science Foundation Graduate Research Foundation program. We thank Ilja Sergueev and Cornelius Strohm for expert help at the ESRF.

**References and Footnotes:**

* corresponding author, e-mail: v.ksenofontov@uni-mainz.de
§ present affiliation: Department of Chemistry, Massachusetts Institute of Technology, 77 Massachusetts Ave, Cambridge, MA 02139 and Department of Chemistry and Department of Physics, The Johns Hopkins University, 3400 N.Charles St, Baltimore, MD 21218, USA

**TABLE 1.** Energy of phonon modes and mode-specific Grüneisen parameters $\gamma_i$ in Fe$_{1.01}$Se. The mode labels (i) are given in Fig. 2 for the 10 K data. Error bars are from fit adjustments or from graphical evaluation (see text). The structures (3) and (7) are not evaluated, only their energy range is given. The Grüneisen parameters $\gamma_i$ are evaluated from the data at 296 K.

| T, K / p, GPa | Energy of phonon modes (i), meV | | | | | | | |
|---|---|---|---|---|---|---|---|---|
| | 1 | 2 | 3 | 4 | 5 | 6 | 7 | 8 |
| 10 K | 5.6(2) | 9.4(2) | 15 - 18 | 20.6(2) | 25.5(2) | 31.5(2) | 33 – 35 | 38.7(2) |
| 66 K | 5.5(2) | 9.4(2) | 15 – 18 | 20.4(3) | 25.5(2) | 31.4(2) | 33 – 35 | 38.6(2) |
| 110 K | 5.5(2) | 9.4(2) | 15 - 18 | 20.3(4) | 25.3(3) | 31.3(3) | 33 – 35 | 38.5(3) |
| 296 K | 5.4(2) | 8.8(2) | 15 - 17 | 20.0(5) | 24.4(3) | 30.1(5) | - | 36.9(4) |
| 296 K / 6.7 GPa | 7.0(2) | 10.0(3) | - | - | 27.5 | 33.1(6) | - | 41.5(6) |
| 296 K | $\gamma_i$ Grüneisen | | | | | | | |
| | 2.4(3) | 1.0(1) | - | - | 0.9(1) | 0.7(1) | - | 0.9(1) |

**TABLE 2.** Lamb-Mössbauer factor $f_{LM}$ and mean force constant $D$ of the Fe sublattice derived from the experimental Fe-partial phonon DOS of Fe$_{1.01}$Se.

| T, K/ p, GPa | 10 K | 66 K | 110 K | 296 K | T = 296 K / 6.7 GPa |
|---|---|---|---|---|---|
| $f_{LM}$ | 0.91(1) | 0.87(1) | 0.83(1) | 0.60(1) | 0.66(1) |
| $D$, N/m | 192(5) | 188(5) | 186(5) | 174(5) | 194(5) |

**TABLE 3.** Average sound velocities $<v_D>$ and corresponding Debye temperatures $\Theta_{D,LT}$ in Fe$_{1.01}$Se derived from the reduced phonon-DOS parameter $\alpha$ (see Equ.1 and Fig. 3). The Debye temperature $\Theta_D$ is calculated from the Lamb-Mössbauer factors $f_{LM}$ of Fe$_{1.01}$Se at 296 K.

| T, K / p, GPa | $\alpha$, meV$^{-3}$ | $<v_D>$, m/s | $\Theta_{D,LT}$, K | $\Theta_D$, K |
|---|---|---|---|---|
| 296 K, ambient pressure | 0.34(2)·10$^{-3}$ | 2.05(4)·10$^3$ | 240(5) | 285(4) |
| 296 K, 6.7 GPa | 0.22(2)·10$^{-3}$ | 2.25(7)·10$^3$ | 278(8) | 317(6) |

**Figures**

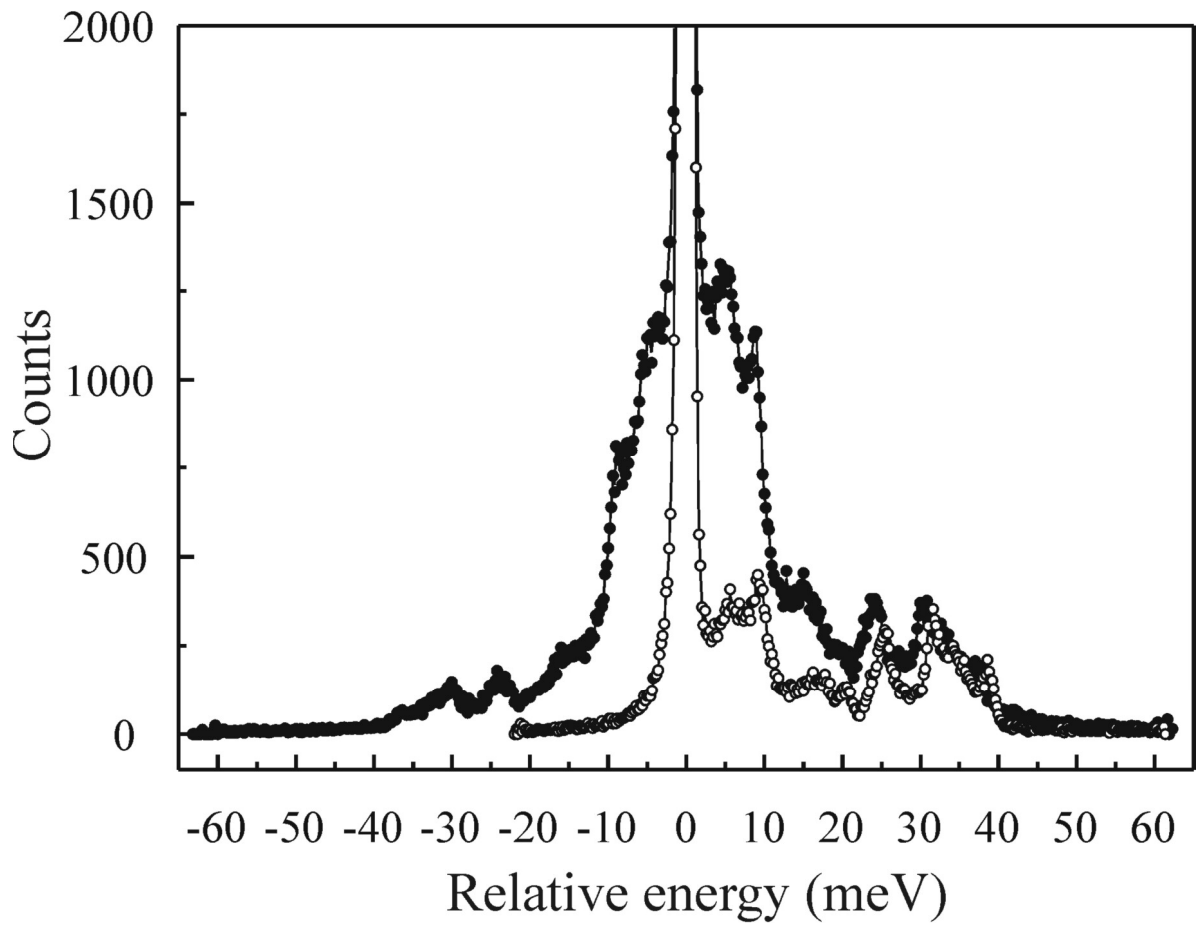

FIG. 1. $^{57}$Fe-NIS spectra of Fe$_{1.01}$Se at room temperature (full circles) and at T = 10 K (open circles).

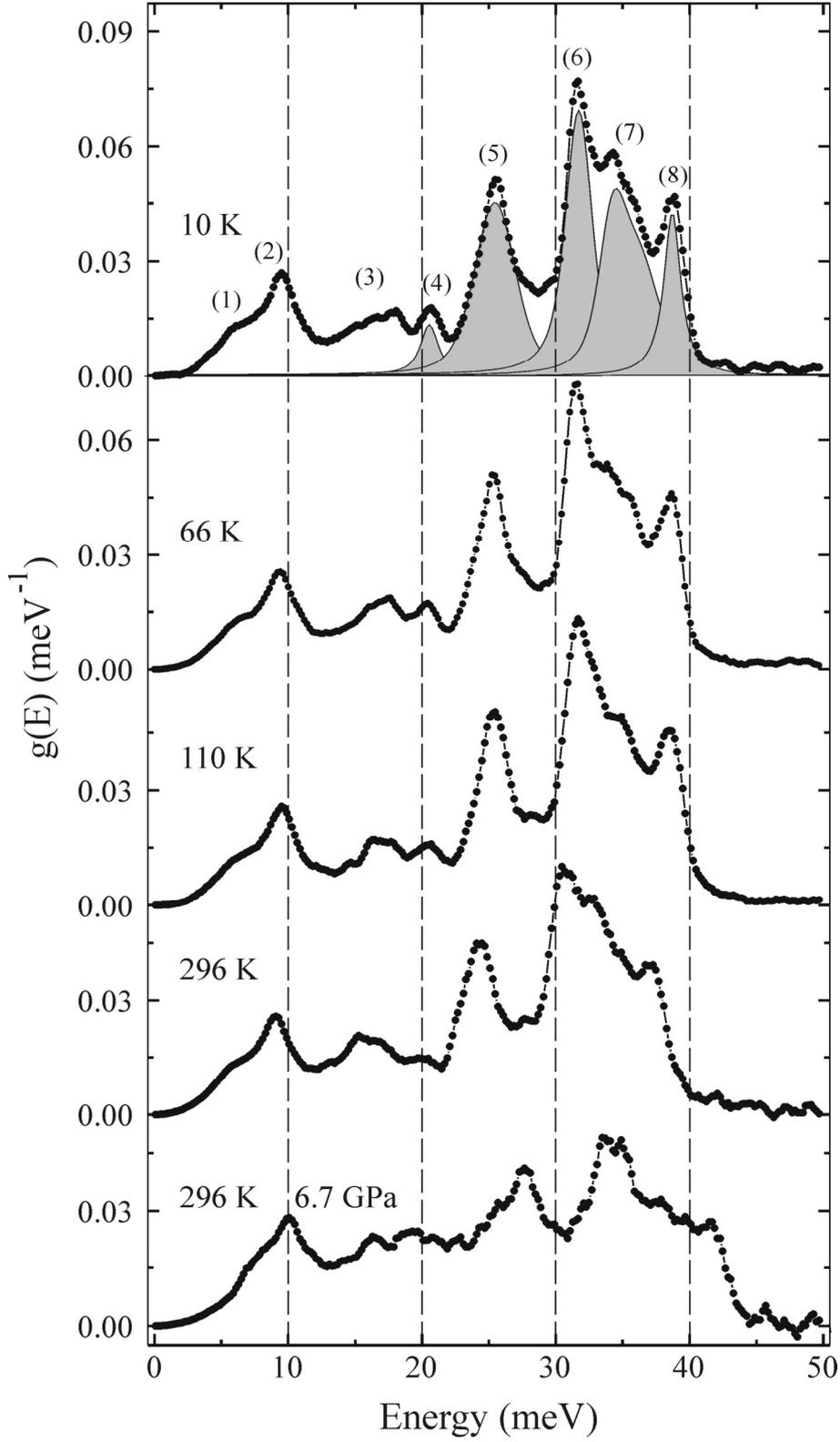

FIG. 2. Fe-partial phonon-DOS $g(E)$ for $Fe_{1.01}Se$ at different temperatures (above) and at a pressure of 6.7 GPa at 296 K (below). At 10 K, resolved optical modes are shaded with light gray.

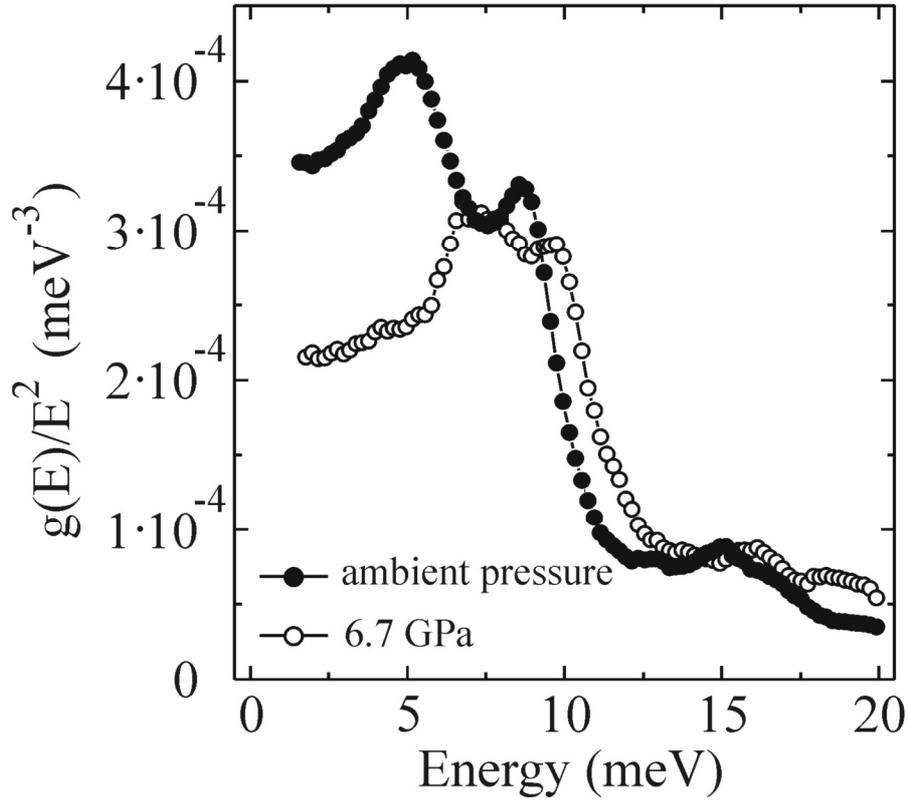

FIG. 3. Fe-partial phonon-DOS for $Fe_{1.01}Se$ at room temperature presented in the reduced form, $g(E)/E^2$, for the low energy range demonstrating the strong modifications of the acoustic modes under pressure. The respective values of $\alpha$ given in Tab. 3 are obtained by extrapolating the data to $E = 0$ meV.